\documentclass[aps,nofootinbib,superscriptaddress,twocolumn,preprintnumbers]{revtex4}

\usepackage{graphicx}
\usepackage{amssymb}
\usepackage{amsmath}
\usepackage{bm}
\usepackage{hyperref}
\usepackage{datetime}
\usepackage{color}
\usepackage[T1]{fontenc} 

\newcommand{\PhT}{\boldsymbol{P}_{hT}}
\newcommand{\kperp}{\boldsymbol{k}_\perp}

\newcommand{\bT}{\boldsymbol{b}_T}

\newcommand{\nslash}{n\kern -0.50em /}
\newcommand{\Sslash}{\kern 0.2 em S\kern -0.50em /}

\begin{document}

\title{Exploring the three-dimensional momentum distribution \\ of longitudinally polarized quarks
  in the proton \\ \vspace{0.2cm}
\normalsize{\textmd{The \textbf{MAP} (Multi-dimensional Analyses of Partonic distributions) Collaboration}}}

\preprint{JLAB-THY-24-4204}

\author{Alessandro Bacchetta}
\thanks{E-mail: alessandro.bacchetta@unipv.it -- \href{https://orcid.org/0000-0002-8824-8355}{ORCID: 0000-0002-8824-8355}}
\affiliation{Dipartimento di Fisica, Universit\`a di Pavia}
\affiliation{INFN Sezione di Pavia, via Bassi 6, I-27100 Pavia, Italy}

\author{Alessia Bongallino}
\thanks{E-mail: alessia.bongallino@ehu.eus --
  \href{https://orcid.org/}{ORCID: 0009-0008-9832-2222}}
\affiliation{Department of Physics \& EHU Quantum Center, University of the Basque Country, Barrio Sarriena s/n, 48940 Leioa, Spain}

\author{Matteo Cerutti}
\thanks{E-mail: mcerutti@jlab.org -- \href{https://orcid.org/0000-0001-7238-5657}{ORCID: 0000-0001-7238-5657}}
\affiliation{Christopher Newport University, Newport News, Virginia 23606, USA}
\affiliation{Jefferson Lab, Newport News, Virginia 23606, USA}

\author{Marco Radici}
\thanks{E-mail: marco.radici@pv.infn.it -- \href{https://orcid.org/0000-0002-4542-9797}{ORCID: 0000-0002-4542-9797}}
\affiliation{INFN Sezione di Pavia, via Bassi 6, I-27100 Pavia, Italy}

\author{Lorenzo Rossi}
\thanks{E-mail: lorenzo.rossi@pv.infn.it -- \href{https://orcid.org/0000-0002-8326-3118}{ORCID: 0000-0002-8326-3118}}
\affiliation{Dipartimento di Fisica, Universit\`a di Pavia}
\affiliation{INFN Sezione di Pavia, via Bassi 6, I-27100 Pavia, Italy}

\begin{abstract}
By analyzing experimental data on semi-inclusive deep inelastic scattering off
longitudinally polarized targets, we extract the
transverse momentum dependence of the quark helicity distribution, i.e., the
difference between the three-dimensional motion of quarks with polarization parallel or antiparallel to the longitudinal polarization of the parent
hadron. We perform the analysis at next-to-leading (NLL) and
next-to-next-to-leading (NNLL) perturbative accuracy. The quality of the
fit is very good for both cases, reaching a $\chi^2$ per number of data points
equal to $1.11$ and $1.09$, respectively.
Although the limited number of data points leads to significant uncertainties,
the data are consistent with an interpretation in which the helicity
distribution is narrower in transverse momentum than the unpolarized
distribution.
\end{abstract}

\maketitle

\section{Introduction}

Quarks are distributed inside the proton in an intricate way,
determined by Quantum
ChromoDynamics (QCD). The reconstruction of the distribution of quarks and
gluons, encoded in the Parton Distribution Functions (PDFs) as a
function of the longitudinal momentum fraction of the parent proton, $x$, is
one of
the most important achievements in the field of hadronic physics and has an
impact on many sectors of physics, from precision studies of the Standard
Model parameters, to the search for physics beyond the
Standard Model, to cosmology~\cite{Gross:2022hyw}.

The distribution of
quarks sharply depends on the orientation of their spins. The clearest example
is provided by the so-called helicity distribution, i.e., the difference in
the distribution of quarks with spin parallel ($q^+$) or antiparallel ($q^-$) to the spin of
the proton when the proton is longitudinally polarized. We will use the
notation $g_1^q(x) = q^+ - q^-$ to denote the helicity distribution of quark $q$ and
$f_1^q(x)= q^+ + q^-$ to denote the unpolarized distribution.
If we interpret PDFs as
parton densitites, their positivity implies that $|g_1^q(x)|\le f_1^q(x)$, which in turn prevents the polarized cross section to become negative (for a more detailed discussion, see Refs.~\cite{deFlorian:2024utd,Candido:2023ujx}).

The helicity distribution has been determined in several works (see
Refs.~\cite{Nocera:2014gqa,Ethier:2020way,Bertone:2024taw,Borsa:2024mss} for
recent extractions). The most salient features presently known can be
summarized as follows: it is large and positive for $u$ quarks, large and
negative for $d$ quarks. The helicity distribution is compatible
with zero for antiquarks, and slightly positive for gluons. The distribution
can be large, and even divergent at low $x$, but it is suppressed relative to the
unpolarized distribution. On the contrary, at large $x$ the helicity
distribution becomes as large as the unpolarized distribution and can saturate
the positivity bound mentioned above.
A detailed knowledge of the helicity distribution is crucial to separate the
various contributions to the proton spin sum rule~\cite{Ji:1996ek},
and can indirectly provide
evidence for the presence of partonic angular momentum (see, e.g.,
Refs.~\cite{Ji:2020ena,Leader:2013jra,Aidala:2012mv} and references therein).

Up to now, the helicity distribution has been studied only as a function of
$x$. But a new frontier in parton distribution studies is the determination
of their dependence on the transverse momentum of quarks, encoded in
Transverse Momentum Dependent PDFs (TMD PDFs). For instance,
we would like to understand if quarks with spin parallel to the proton spin tend to
have smaller or larger transverse momentum than quarks with antiparallel spin. The goal of the present work is to answer such kind of
questions through the determination of the helicity TMD PDF from existing data.\footnote{The helicity TMD PDF has been often
denoted in
the literature as $g_{1L}$, but for convenience in this present work we will
denote it as its collinear counterpart, $g_1$.}
In our analysis, we use data measured by the HERMES
Collaboration~\cite{HERMES:2018awh} in Semi-Inclusive Deep Inelastic
Scattering (SIDIS) with longitudinally polarized targets.
Recent studies of other polarized TMD PDFs can be found in Refs.~\cite{Cammarota:2020qcw,Bacchetta:2020gko,Echevarria:2020hpy,Bury:2020vhj,Boglione:2021aha} for the Sivers function $f_{1T}^\perp$, in Refs.~\cite{Kang:2015msa,Cammarota:2020qcw,Gamberg:2022kdb,Boglione:2024dal} for the transversity $h_1$, in Ref.~\cite{Bhattacharya:2021twu} for the worm-gear $g_{1T}$, and in Ref.~\cite{Lefky:2014eia} for the pretzelosity $h_{1T}^\perp$.

For a consistent determination of any polarized TMD PDF, it is essential to start
from a reliable extraction of the unpolarized TMD PDF and, when SIDIS data
are involved, also of unpolarized TMD Fragmentation Functions (TMD FFs).
Recent extractions of the unpolarized TMD PDF $f_1$ have been published in
Refs.~\cite{Bacchetta:2017gcc,Scimemi:2017etj,Bertone:2019nxa,Scimemi:2019cmh,Bacchetta:2019sam,Bury:2022czx,Bacchetta:2022awv,Moos:2023yfa,Bacchetta:2024qre}. The high level of perturbative accuracy reached in the theoretical framework and the large data sets analyzed with a very good quality of the fits are such that TMD PDFs have entered a precision era where they can potentially impact precision studies of Standard
Model parameters like, e.g., the determination of the $W$ mass at hadron colliders~\cite{Bacchetta:2018lna}.
Recent extraction of the unpolarized TMD FF $D_1$ are presented in
Refs.~\cite{Bacchetta:2017gcc,Scimemi:2019cmh,Bacchetta:2022awv,Bacchetta:2024qre}. In
this work, we choose to start from the MAPTMD22
extraction~\cite{Bacchetta:2022awv} and we analyze the helicity TMD PDF with
exactly the same formalism and computational framework~\cite{NangaParbat}.
We reach the
next-to-next-to-leading (NNLL) level of logarithmic accuracy in resumming the soft gluon radiation, the highest
presently achievable, but we discuss results also at NLL accuracy.

Our work is particularly relevant also because one of the main goals of existing and
planned experimental
facilities is to provide data to improve the determinations of the spin
dependence of
multi-dimensional quark distributions, including the helicity distribution~\cite{Boer:2011fh,Accardi:2012qut,Dudek:2012vr,Aschenauer:2016our,Accardi:2023chb}.

When this work was close to completion, an extraction of the helicity TMD PDF
appeared in Ref.~\cite{Yang:2024drd}. We briefly discuss the differences with this recent
work at the end of Sec.~\ref{s:results}.

\section{Formalism}

For the SIDIS process $\vec{e} + \vec{p} \to e' + h + X$ involving the scattering of an electron and a proton, both longitudinally polarized, and the inclusive production of a hadron with fractional energy $z$ and transverse momentum $\PhT$, the starting point of our analysis is the double spin
asymmetry $A_1$. It is normally defined for inclusive DIS~\cite{Diehl:2005pc,Filippone:2001ux}, but it can be generalized to
the SIDIS case at low transverse momentum using TMD
factorization~\cite{Ji:2004xq,Collins:2011zzd}:
\begin{widetext}
\begin{align}
    A_1(x,z,Q,|\PhT|) &= \frac{d\sigma^{\to \leftarrow} - d\sigma^{\to \to} + d\sigma^{\leftarrow \to} - d\sigma^{\leftarrow \leftarrow}}{d\sigma^{\to \leftarrow} + d\sigma^{\to \to} + d\sigma^{\leftarrow \to} + d\sigma^{\leftarrow \leftarrow}} \notag \\
    &= \frac{
        \displaystyle \sum_{a=q,\bar{q}} e_a^2
        \int_0^{+\infty} d|\bT| |\bT|  \, J_0\left(\frac{|\bT||\PhT|}{z}\right)
        \hat{g}_{1}^a(x,\bT^2,Q)
        \hat{D}_1^{a\rightarrow h} (z,\bT^2,Q)
    }{
        \displaystyle \sum_{a=q,\bar{q}} e_a^2
        \int_0^{+\infty} d|\bT| |\bT| \, J_0\left(\frac{|\bT||\PhT|}{z}\right)
        \hat{f}_{1}^a(x,\bT^2,Q)
        \hat{D}_1^{a\rightarrow h} (z,\bT^2,Q)
    } \, ,
\label{e:A1}
\end{align}
where the invariant mass $Q$ of the exchanged virtual photon is the hard scale
of the SIDIS process, and $J_0$ is a Bessel function. The formula
neglects power corrections of the type $\PhT^2/z^2 Q^2$ or $M^2/Q^2$.
The denominator of Eq.~\eqref{e:A1} corresponds to the unpolarized cross
section and involves the unpolarized TMD PDF $\hat{f}_1$ and TMD FF
$\hat{D}_1$ in Fourier-conjugate $\bT$-space, where the factorization formula is simpler. In principle, these functions should depend on two scales that usually are identified with the hard scale $Q$. In a similar way, the numerator of Eq.~\eqref{e:A1} involves the helicity TMD PDF $\hat{g}_1$.

Using the same prescriptions as in the MAPTMD22
analysis~\cite{Bacchetta:2022awv}, which are essentially based on the
so-called Collins--Soper--Sterman (CSS)
approach~\cite{Collins:1984kg,Collins:2011zzd},  
the TMD PDFs $\hat{f}_1$ and $\hat{g}_1$
evolved from an arbitrary low scale $Q_0$ to the hard scale $Q$ can be written
as
\begin{align}
\hat{f}_1(x,\bT^2,Q) &= \left[ C^f \otimes f_1 \right](x,
b_\ast(\bT^2)) \, f_{\text{NP}}(x,\bT^2,Q_0) \,  e^{S (\mu_{b_\ast}^2, Q^2)} \ e^{g_K(\bT^2) \ln (Q^2 / Q_0^2)} \, ,
\label{e:f1evo} \\
\hat{g}_{1}(x,\bT^2,Q) &= \left[ C^g \otimes g_1 \right](x, b_\ast(\bT^2))
\, g_{\text{NP}}(x,\bT^2,Q_0) \,
\,  e^{S (\mu_{b_\ast}^2, Q^2)} \ e^{g_K(\bT^2) \ln (Q^2 / Q_0^2)} \, .
\label{e:g1evo}
\end{align}
\end{widetext}
A similar formula holds also for the TMD FF $\hat{D}_1$.
Using the Operator Product Expansion (OPE), the TMD PDFs $\hat{f}_1$, $\hat{g}_1$ can be matched onto the corresponding PDFs $f_1$ and $g_1$ through the Wilson coefficients $C^f$ and $C^g$, respectively. Both $C^f$, $C^g$ and the evolution operator $S$ can be calculated perturbatively. In this work, we reach the NLL and NNLL perturbative accuracy, according to Tab.~1 of Ref.~\cite{Bacchetta:2019sam}. Note that for the numerator of Eq.~\eqref{e:A1} the NNLL level is the highest possible accuracy because the Wilson coefficients $C^g$ are known only up to order
${\cal O} (\alpha_s)$~\cite{Gutierrez-Reyes:2017glx}.

In Eqs.~\eqref{e:f1evo} and \eqref{e:g1evo}, $f_{\text{NP}}$,
$g_{\text{NP}}$, $g_K$ are arbitrary nonperturbative functions that must be
parametrized to experimental data. The scale
$\mu_{b_\ast}$ is defined as  $\mu_{b_\ast}=  2 e^{-\gamma_E}/b_\ast$,
where $\gamma_E$ is the Euler
constant, and $b_\ast$ is chosen according to an appropriate
prescription.
We take $f_{\text{NP}}$, $g_K$ and $b_\ast$
from the MAPTMD22 analysis~\cite{Bacchetta:2022awv} (and similarly for the $D_{\text{NP}}$ of the TMD FF $\hat{D}_1$).
We model the nonperturbative $g_{\text{NP}}$ in momentum space as the product of the nonperturbative $f_{\text{NP}}$ and a Gaussian with an $x$-dependent width:
\begin{equation}
\label{e:g1_NP}
g_{\text{NP}}(x, \kperp^2, Q_0) = \dfrac{f^{\text{MAP22}}_{\text{NP}}(x, \kperp^2, Q_0) \, e^{-\frac{\kperp^2}{w_1(x)}}} {k_{\text{norm}}(x)} \, ,
\end{equation}
where $\kperp$ is the transverse momentum of quarks with respect to the proton
momentum direction, and $k_{\text{norm}}(x)$ depends on $w_1 (x)$ and ensures
that the integration over $\kperp$ of $g_{\text{NP}}$ equals unity (for
convenience, the expression of $k_{\text{norm}}(x)$ is reproduced in Eq.~\eqref{e:k_norm}).
The Gaussian width $w_1(x)$ has a crucial role in granting the positivity constraint $|g_1| \leq f_1$ also at the TMD level. 
In fact, by taking the ratio between the expressions in momentum space of Eq.~\eqref{e:g1evo} and Eq.~\eqref{e:f1evo} at the NLL level and at the initial scale $Q_0 = 1$ GeV, because of Eq.~\eqref{e:g1_NP} we get
\begin{equation}
\label{e:positivity}
\frac{g_{1}(x, \kperp^2, Q_0)}{f_{1}(x, \kperp^2, Q_0)} = \frac{g_{1}(x, Q_0)}{f_1(x, Q_0)}  \frac{ e^{- \frac{\kperp^2}{w_1(x)}}}{k_{\text{norm}}(x)} \, .
\end{equation}
For $|\kperp| \to 0$, the term $e^{-\kperp^2 / w_1(x)} / k_{\text{norm}}(x)$ could potentially become very large, thus causing a possible violation of the positivity constraint unless we impose 
\begin{equation}
\label{e:g1pos}
\frac{g_{1}(x, Q_0)}{f_{1}(x, Q_0)} \frac{1}{k_{\text{norm}}(x)} \leq 1 \, .
\end{equation}
This condition implies that $w_1(x)$ must be bounded from below. Therefore, we choose
\begin{equation}
\label{e:w1}
w_{1}(x) = f_{\text{pos.}}(x) + N_{1g}^2 \frac{ (1-x)^{\alpha_{1g}^2} x^{\sigma_{1g}}}{ (1-\hat{x})^{\alpha_{1g}^2}\hat{x}^{\sigma_{1g}}} \, ,
\end{equation}
where $N_{1g}, \, \alpha_{1g}, \, \sigma_{1g}$ are free parameters and $\hat{x} = 0.1$. The function $f_{\text{pos.}}(x)$ depends on the
MAPTMD22 parameters, and it can be conveniently approximated as in Eq.~\eqref{e:f_pos}. 
Eq.~\eqref{e:g1pos} is
maintained for all values of $x$ in the analyzed range $[10^{-4},0.7]$. It
also holds for higher values of $x$ if higher-order and target-mass
corrections are neglected. However, since these corrections become significant
in that region, we limit our analysis to the specified range
$10^{-4} \leq x \leq 0.7$.

\section{Results}
\label{s:results}

In this section, we present the key result of this study, the extraction of
the helicity TMD PDF from a fit to SIDIS experimental data for the double spin
asymmetry $A_1$ of Eq.~\eqref{e:A1}.
We use data for positive and negative charged pion/kaon
production from deuterium and proton targets from the \textsc{HERMES}
Collaboration~\cite{HERMES:2018awh}.

We apply the same kinematic cuts of the
MAPTMD22 analysis~\cite{Bacchetta:2022awv} in order to keep consistency with
the unpolarized TMDs entering the denominator of $A_1$ and, more importantly,
to fulfill the conditions for TMD factorization. Therefore, we do not include data on deuteron target from the \textsc{COMPASS} Collaboration (see Fig.~6 of Ref.~\cite{COMPASS:2016klq}) nor data from the \textsc{CLAS6} Collaboration~\cite{CLAS:2017yrm} because they are not compatible with our kinematic cuts.

In total, we fit 291 data points. Our error analysis is performed with the so-called bootstrap method, namely by fitting an ensemble of Monte Carlo (MC) replicas of the experimental data. As in previous works of the \textsc{MAP} Collaboration~\cite{Bacchetta:2022awv,Cerutti:2022lmb,Bacchetta:2024qre}, we consider
the $\chi^2$ value of the best fit to the unfluctuated data as the most representative indicator of the quality of the fit. The data set has statistical and systematic uncertainties. We consider the former as uncorrelated while the latter as fully correlated. The expression of the $\chi^2$ contains a penalty term due to correlated uncertainties, that is described through nuisance parameters determined by minimizing the full $\chi^2$ on data (for more details see Refs.~\cite{Bacchetta:2022awv,Cerutti:2022lmb,Bacchetta:2024qre}).

For the denominator of the asymmetry $A_1$ in Eq.~\eqref{e:A1}, we take the unpolarized TMD PDF $\hat{f}_1$ and TMD FF $\hat{D}_1$ from the MAPTMD22 extraction~\cite{Bacchetta:2022awv}. For the collinear polarized PDFs in Eq.~\eqref{e:g1evo}, we choose the NNPDFpol1.1 set~\cite{Nocera:2014gqa} with respect to more recent extractions~\cite{Bertone:2024taw,Borsa:2024mss} because it includes parametrizations of $g_1$ also  at lower perturbative order, that are needed for our analysis at NLL and NNLL accuracy. The NNPDFpol1.1 set contains 100 MC members. We generate the same number of replicas of the $A_1$ data points, we fit them, and we associate the $i$--th replica of the helicity PDF and the corresponding extracted helicity TMD PDF to the same replica of the unpolarized TMDs in the MAPTMD22 extraction. In this way, we propagate the uncertainty in the extraction of helicity PDFs onto the uncertainty of helicity TMD PDFs.

We perform our analysis at NLL and NNLL perturbative accuracy. The quality of the fit for both accuracies is shown in Table~\ref{t:chi2_g1}, where the $\chi^2$ per number of data points $N_{\rm dat}$ are listed for each considered experimental data set.

\begin{table}[h]
    \centering
    \begin{tabular}{|l|c|c|c|}
    \hline
         Experiment & $N_{\text{dat}}$  & $\chi_{\text{NLL}}^2 / N_{\text{dat}}$ & $\chi_{\text{NNLL}}^2 /N_{\text{dat}}$ \\ \hline \hline
         \textsc{HERMES} ($d \rightarrow \pi^+$) & 47 &  1.34 & 1.30 \\ \hline
         \textsc{HERMES} ($d \rightarrow \pi^-$) & 47 &  1.10 & 1.08 \\ \hline
         \textsc{HERMES} ($d \rightarrow K^+$) & 46 & 1.26 & 1.25 \\ \hline
         \textsc{HERMES} ($d \rightarrow K^-$) & 45 & 0.93 & 0.89 \\ \hline
         \textsc{HERMES} ($p \rightarrow \pi^+$) & 53 &  1.17 & 1.21 \\ \hline
         \textsc{HERMES} ($p \rightarrow \pi^-$) & 53 &  0.86 & 0.86 \\ \hline \hline
         Total & 291 & 1.11  & 1.09 \\ \hline
    \end{tabular}
    \caption{Breakdown of $\chi^2$ per number of data points $N_{\text{dat}}$ for the best fits of  \textsc{HERMES} data~\cite{HERMES:2018awh} of double spin asymmetry $A_1$ in Eq.~\eqref{e:A1} at NLL and NNLL accuracy.}
    \label{t:chi2_g1}
\end{table}

We note that the global quality of the fit slightly increases at higher accuracy. We also observe that the $\chi^2$ on the experimental data for the $\pi^+$ production are larger than for other fragmentation channels. The same feature was observed in the MAPTMD22 extraction of unpolarized quark TMDs~\cite{Bacchetta:2022awv} and it is most likely related to the smaller experimental uncertainties in this fragmentation channel. In the Supplemental Material, we show the comparison between experimental data and results of the fit for all kinematic bins.

\begin{table}[h]
    \centering
    \begin{tabular}{|c|c|c|c|}
    \hline
    Parameters & $N_{1g}$ & $\alpha_{1g}$ & $\sigma_{1g}$ \\ \hline \hline
    NLL & $0.70 \pm 0.54$ & $27.81 \pm 27.70$ & $0.42 \pm 0.86$ \\ \hline
    NNLL & $0.87 \pm 0.72$ & $6.73 \pm 6.58$  & $3.04 \pm 3.09$  \\ \hline
    \end{tabular}
    \caption{Average values and uncertainties (68\% C.L.) of the free parameters in the helicity TMD PDF at NLL and
      NNLL accuracy.}
    \label{t:params_g1}
\end{table}

In Tab.~\ref{t:params_g1}, we show the mean average values and associated errors at 68\% confidence level (C.L.)  for the free parameters in Eq.~\eqref{e:w1} at both NLL and NNLL accuracy. We note that in both cases the free parameters are poorly constrained. This is a consequence of the small number of available experimental data.

\begin{figure}[h]
\centering
\includegraphics[width=0.48\textwidth]{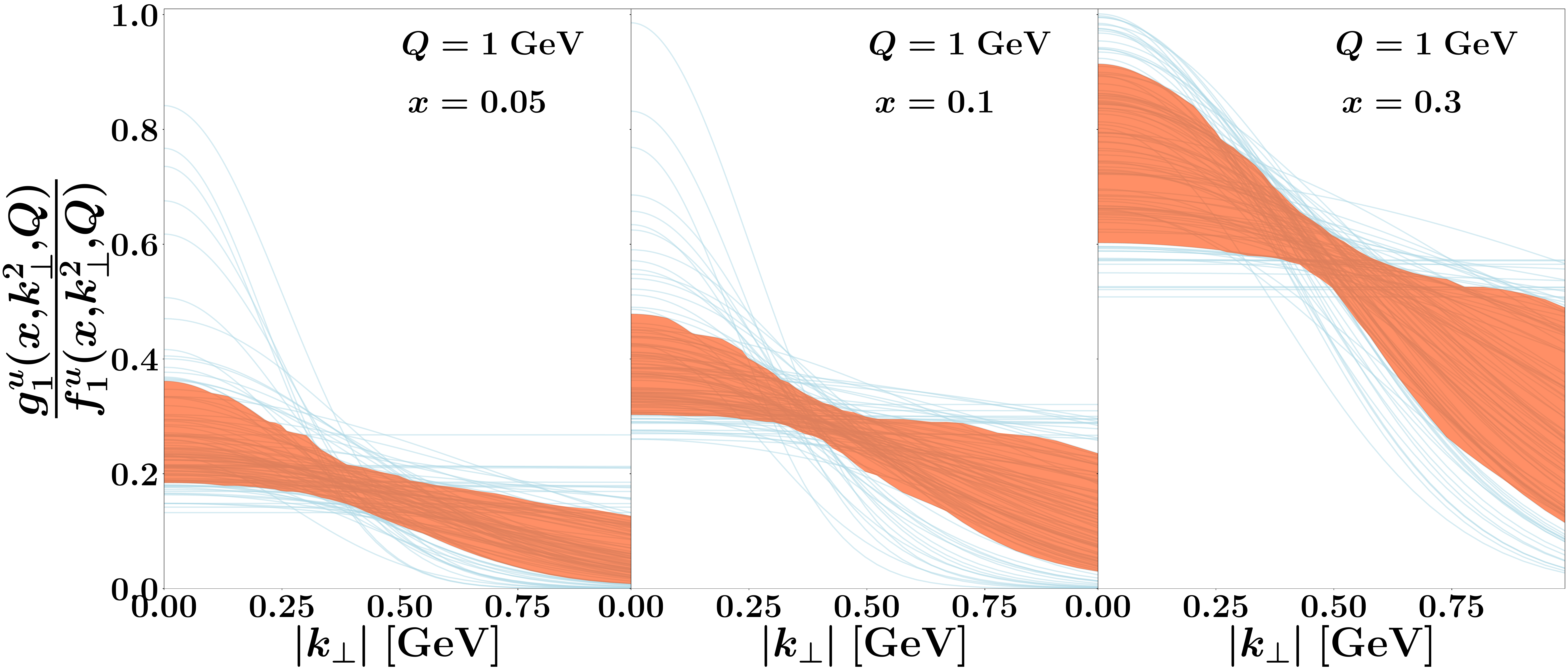}
\caption{Ratio between the helicity and unpolarized TMD PDFs for quark $u$ in a proton at NNLL as a function of the quark transverse momentum $|\kperp|$, at $Q = 1$ GeV and for $x = 0.05$ (left panel), $x = 0.1$ (central panel), and $x = 0.3$ (right panel). Light blue lines for all replicas of the fit, orange band represents the 68\% C.L.}
\label{f:g1/f1}
\end{figure}

In Fig.~\ref{f:g1/f1}, we show the ratio between the helicity and the
unpolarized TMD PDFs for a quark $u$ in a proton at NNLL, as a function of the
quark transverse momentum $|\kperp|$ at $Q = 1$ GeV and for $x = 0.05$ (left
panel), $x = 0.1$ (central panel), and $x = 0.3$ (right panel). The light blue
lines represent all the replicas of our fit, while the orange band includes
only 68\% of them, obtained by excluding for each bin the largest and smallest
16\% of them. We note that  the $|\kperp|$ distribution of the ratio depends on $x$. At relatively small $x$, the ratio is almost flat: the shape of the $|\kperp|$ distribution of helicity and unpolarized TMD PDFs is approximately the same, the former being just rescaled from the latter. At larger $x$, the trend is different showing that the helicity TMD PDF has a sharper $|\kperp|$ distribution than the unpolarized one.

\begin{figure}[h]
\centering
\includegraphics[width=0.45\textwidth]{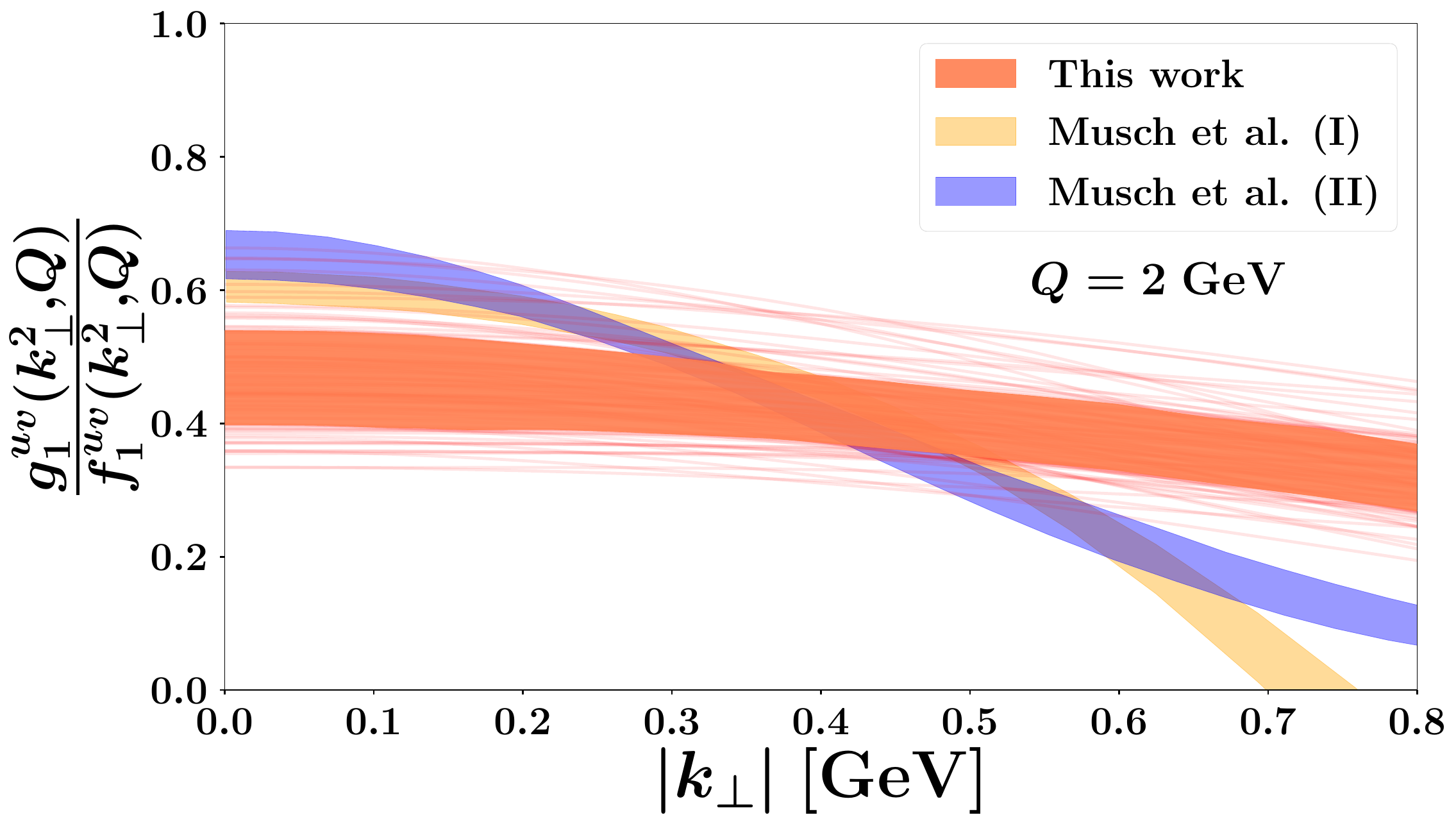}
\caption{Ratio between the helicity and unpolarized TMD PDFs for valence $u_v$ quark integrated upon $x$, as a function of $|\kperp|$ at $Q = 2$ GeV.  Orange lines for all replicas of this fit at NNLL (orange band for the 68\% C.L.), yellow and blue bands for the lattice calculations of Ref.~\cite{Musch:2010ka}.}
\label{f:lattice}
\end{figure}

In the last years, efforts have been made to compute TMDs with lattice QCD
(see Ref.~\cite{Constantinou:2020hdm} and references therein).
In order to make a comparison between our phenomenological extraction and lattice calculations, in Fig.~\ref{f:lattice} we show the ratio between helicity and unpolarized TMD PDFs for the valence quark $u_v$, integrated over $x$ as a function of $|\kperp|$ at $Q = 2$ GeV. The orange lines represent all the replicas of our fit at NNLL (orange band for the 68\% C.L.). The yellow and blue bands show the results of the lattice calculation of Ref.~\cite{Musch:2010ka}.
All the curves from our extraction are obtained by a numerical integration in the range $10^{-3} \leq x \leq 0.9$ in order to avoid numerical issues at the endpoints.

We observe that the result of our phenomenological extraction is in fair
agreement with the lattice calculation
but have a milder slope.
A similar result is obtained at NLL accuracy. Further studies on the comparison with lattice results are certainly needed to assess the compatibility between the two different approaches.

While approaching the completion of this work, another extraction of the
helicity TMD PDF appeared in Ref.~\cite{Yang:2024drd}. The main differences
with our extraction are twofold. First, the authors of
Ref.~\cite{Yang:2024drd} implement the scale dependence of TMDs using
the $\zeta$-prescription~\cite{Scimemi:2018xaf} rather than the
CSS approach~\cite{Collins:2011zzd} used in
Eqs.~\eqref{e:f1evo} and \eqref{e:g1evo}. Second, they include in the fit the 
\textsc{CLAS6} experimental data and apply a more
conservative transverse momentum cut to \textsc{HERMES} data, resulting in  
a smaller total number of analyzed data points compared to our work.
More
importantly, the authors of Ref.~\cite{Yang:2024drd} modify the $x$ dependence
of the collinear helicity PDF.
This has two important consequences: it breaks the matching
between helicity PDF and TMD PDF in the OPE formula, and it implies that the
integral over $\kperp$ of the helicity TMD PDF does not reproduce the helicity
PDF, even at NLL. As a final remark, the positivity constraint $|g_1|\le f_1$
is not enforced on the extracted helicity TMD PDF, and indeed it appears to be
violated.

\section{Conclusions}

We have extracted the helicity TMD PDF of quarks from \textsc{HERMES}
data~\cite{HERMES:2018awh} for the double spin asymmetry in SIDIS kinematic
conditions that satisfy the TMD factorization requirement. We evaluate the
denominator of the asymmetry by using the unpolarized TMDs of the MAPTMD22
extraction~\cite{Bacchetta:2022awv}, and we consistently describe the helicity
TMD PDF in the same framework reaching the NNLL perturbative accuracy, which
is the current highest possible accuracy. We have parametrized the
nonperturbative part of the helicity TMD PDF such that the positivity
constraint is fulfilled by construction. Consistently with the MAPTMD22
extraction, we have performed the error analysis using the bootstrap method
and correcting the $\chi^2$ expression with a penalty for correlated
uncertainties.
At large quark transverse momenta, the helicity TMD PDF can be matched onto the collinear helicity PDF through OPE. For the collinear helicity PDF, we have used the NNPDFpol1.1 set~\cite{Nocera:2014gqa}, and we have propagated the PDF uncertainties onto the TMD PDF by selecting a one-to-one correspondence between the replicas of the former and of the latter.

The quality of the fit is very good for both NLL and NNLL accuracies, reaching a $\chi^2$ per number of data points equal to $1.11$ and $1.09$, respectively. The shape of the transverse momentum distribution of the quark helicity TMD PDF shows a dependence on the longitudinal fractional momentum $x$ and tends to deviate from the one of the unpolarized TMD PDF at large $x$. Also, it nicely compares with the only lattice calculation available so far, after integrating upon $x$.

The nonperturbative part of the helicity TMD PDF is described by three free parameters which are poorly constrained because of the limited size of the experimental data set. New data from Jefferson Lab, as well as from the Electron-Ion Collider, will be important to refine the model of the arbitrary nonperturbative part of the helicity TMD PDF and, ultimately, to get more precise information regarding the spin content of the nucleon.

\begin{acknowledgments}
We thank Valerio Bertone for his help in the early stages of this work.
This work is supported by by the European Union ``Next
Generation EU'' program through the Italian PRIN 2022 grant
n.~20225ZHA7W and by the European Union's Horizon 2020 programme
under grant agreement No.~824093 (STRONG2020).
This material is also based upon work supported by the U.S. Department of Energy, Office of Science, Office of Nuclear Physics under contract DE-AC05-06OR23177.
\end{acknowledgments}

\newpage

\bibliography{g1TMD}
\bibliographystyle{myrevtex}

\appendix

\begin{widetext}

  \section{Useful explicit formulas}

We present here the expressions of all components that constitute the
nonperturbative part of the helicity TMD PDF in
Eq.~\eqref{e:g1_NP}. We begin by detailing the nonperturbative part employed
in the MAPTMD22 analysis in $\kperp$ space: 
\begin{equation}
    f_{\text{NP}}^{\text{MAP22}}(x,\kperp^2,Q_0) = \frac{\exp\left(-\frac{\kperp^2}{g_{1A}(x)}\right) + \kperp^2 \lambda^2 \exp\left(-\frac{\kperp^2}{g_{1B}(x)}\right) + \lambda_2^2 \exp\left(-\frac{\kperp^2}{g_{1C}(x)}\right)}{\pi \left(g_{1A}(x) + \lambda^2 g_{1B}(x)^2 + \lambda_2^2 \, g_{1C}(x)\right)} \, ,
\end{equation}
where the $x$-dependent gaussian widths are defined as
\begin{equation}
    g_{\{1A,1B,1C\}}(x) = N_{\{1,2,3\}} \frac{(1-x)^{\alpha^2_{\{1,2,3\}}} \, x^{\sigma_{\{1,2,3\}}}}{(1-\hat{x})^{\alpha^2_{\{1,2,3\}}} \, \hat{x}^{\sigma_{\{1,2,3\}}}} \, ,
\end{equation}
with $\hat{x}=0.1$. The $N_i, 
, \alpha_i, \, \sigma_i$ ($i = 1,2,3$) and $\lambda_j$ ($j = 1,2$) are the parameters extracted at NLL and NNLL in the MAPTMD22 work~\cite{Bacchetta:2022awv} and they are available in the public {\tt NangaParbat} repository.

The factor $ k_{\text{norm}}(x) $, introduced to ensure the normalization of the non-perturbative part, is given by the following expression:
\begin{equation}
\label{e:k_norm}
    k_{\text{norm}}(x) = w_{1}(x) \frac{
    \frac{g_{1A}(x)}{g_{1A}(x) + w_{1}(x)} + \lambda^2 \frac{g^2_{1B}(x) w_{1}(x)}{(g_{1B}(x) + w_{1}(x))^2} + \lambda_2^2 \frac{g_{1C}(x)}{g_{1C}(x) + w_{1}(x)}
    }{
    g_{1A}(x) + \lambda^2 g_{1B}^2(x) + \lambda_2^2 g_{1C}(x)
    }.
\end{equation}
This function depends on the MAPTMD22 parameters and on the function $w_{1}(x)$
defined in Eq.~\eqref{e:w1}, which we reproduce here for convenience
\begin{equation}
\label{e:w1_rep}
w_{1}(x) = f_{\text{pos.}}(x) + N_{1g} \frac{ (1-x)^{\alpha_{1g}^2} x^{\sigma_{1g}}}{ (1-\hat{x})^{\alpha_{1g}^2}\hat{x}^{\sigma_{1g}}} \, .
\end{equation}

The function $f_{\text{pos.}}(x)$ is chosen as a simple
analytical formula 
\begin{equation}
f_{\text{pos.}}(x) \approx c + h^2 e^{-\frac{(x-\mu)^2}{\sigma^2}} \, ,
\label{e:f_pos}
\end{equation}
where the values of the parameters $c$, $h^2$, $\mu$, and $\sigma$ will be
available in the public {\tt NangaParbat}
repository~\cite{NangaParbat}.

\section{Data/theory comparison plots}

We report here the comparison of double spin asymmetries $A_1$ between our analysis and experimental data for all kinematic bins. The red lines all the MC replicas of the theoretical predictions.
\begin{figure}[h]
\centering
\includegraphics[width=0.7\textwidth]{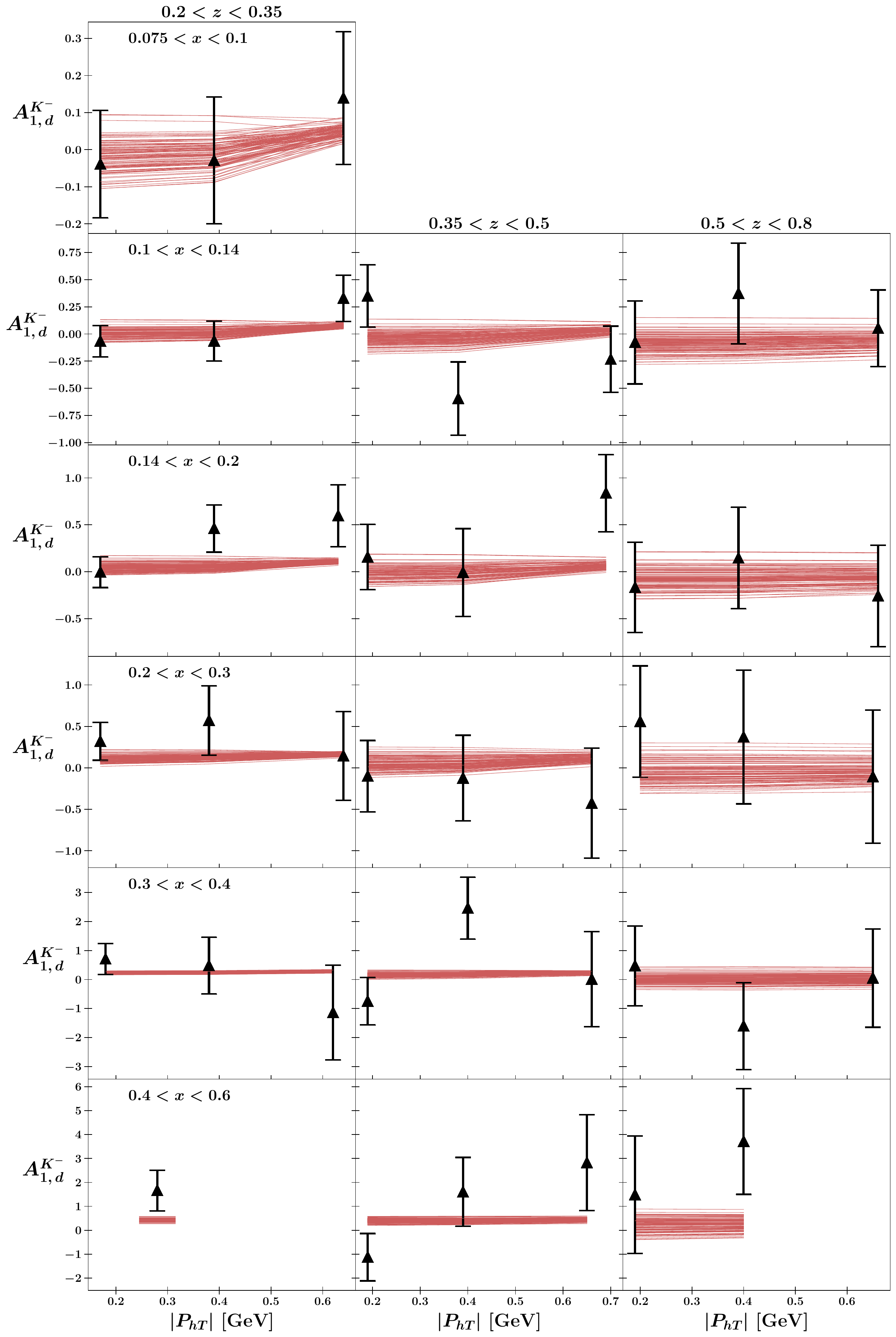}
\caption{Comparison between \textsc{HERMES} asymmetry data and theoretical predictions (red lines) for the production of negative kaon off a deuteron target for a selection of $x$ and $z$ bins. The results are reported as a function of the transverse momentum $|\PhT|$ of the final-state hadron.}
\label{f:A1_d_km}
\end{figure}

\begin{figure}[h]
\centering
\includegraphics[width=0.7\textwidth]{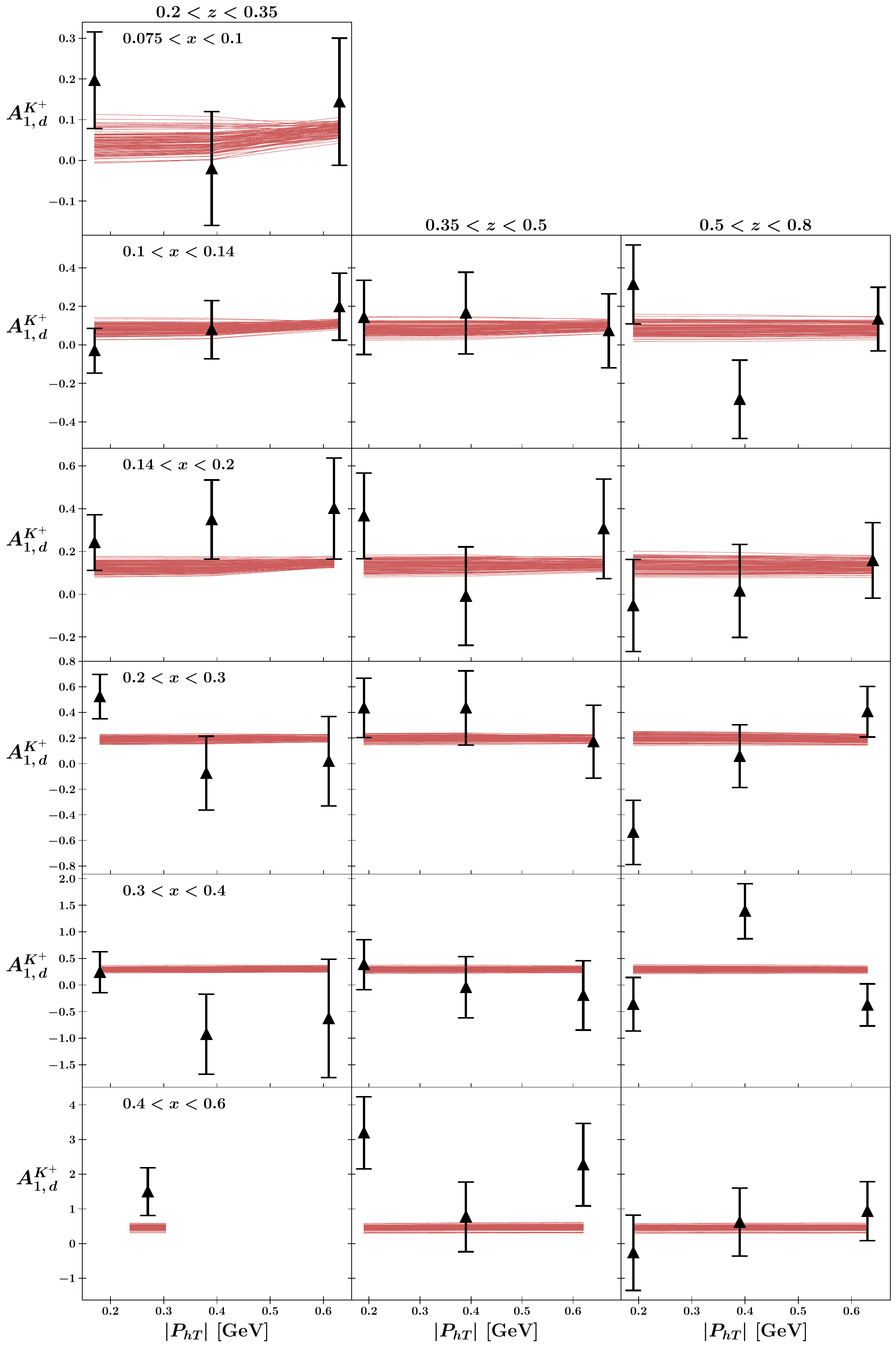}
\caption{Comparison between \textsc{HERMES} asymmetry data and theoretical predictions (red lines)  for the production of positive kaon off a deuteron target for a selection of $x$ and $z$ bins. The results are reported as a function of the transverse momentum $|\PhT|$ of the final-state hadron.}
\label{f:A1_d_kp}
\end{figure}

\begin{figure}[h]
\centering
\includegraphics[width=0.7\textwidth]{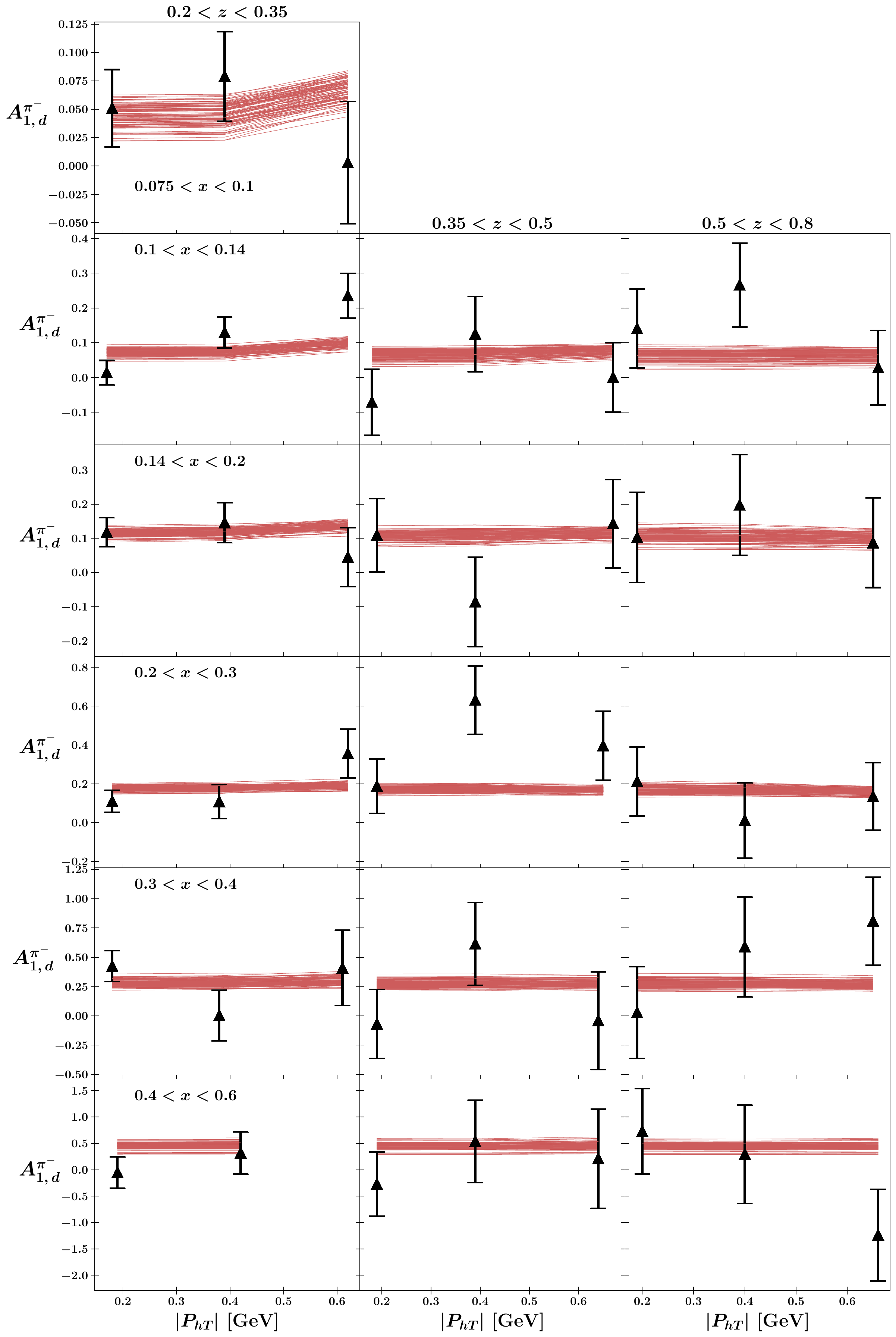}
\caption{Comparison between \textsc{HERMES} asymmetry data and theoretical predictions (red lines)  for the production of negative pion off a deuteron target for a selection of $x$ and $z$ bins. The results are reported as a function of the transverse momentum $|\PhT|$ of the final-state hadron.}
\label{f:A1_d_pim}
\end{figure}

\begin{figure}[h]
\centering
\includegraphics[width=0.7\textwidth]{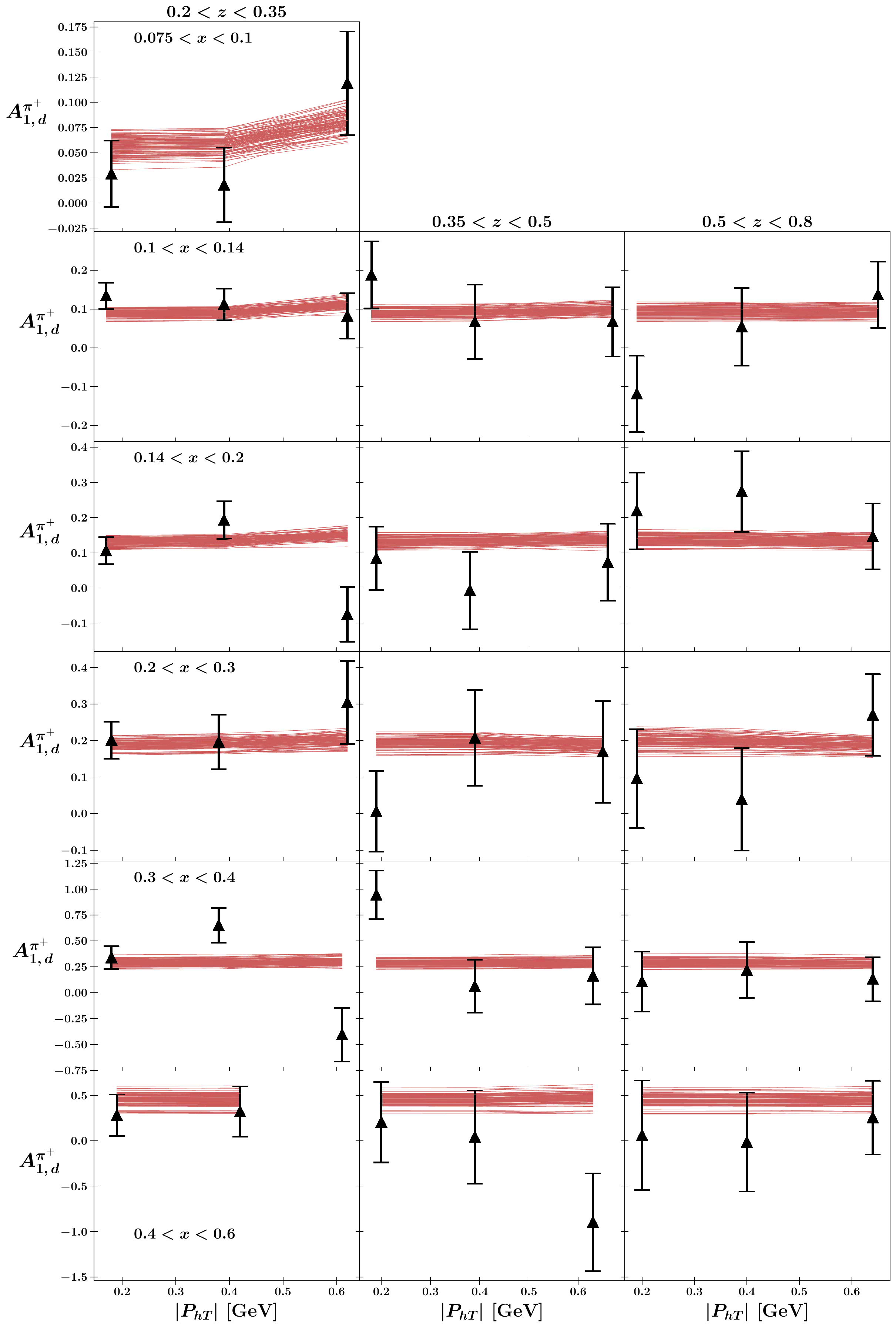}
\caption{Comparison between \textsc{HERMES} asymmetry data and theoretical predictions (red lines)  for the production of positive pion off a deuteron target for a selection of $x$ and $z$ bins. The results are reported as a function of the transverse momentum $|\PhT|$ of the final-state hadron.}
\label{f:A1_d_pip}
\end{figure}

\begin{figure}[h]
\centering
\includegraphics[width=0.7\textwidth]{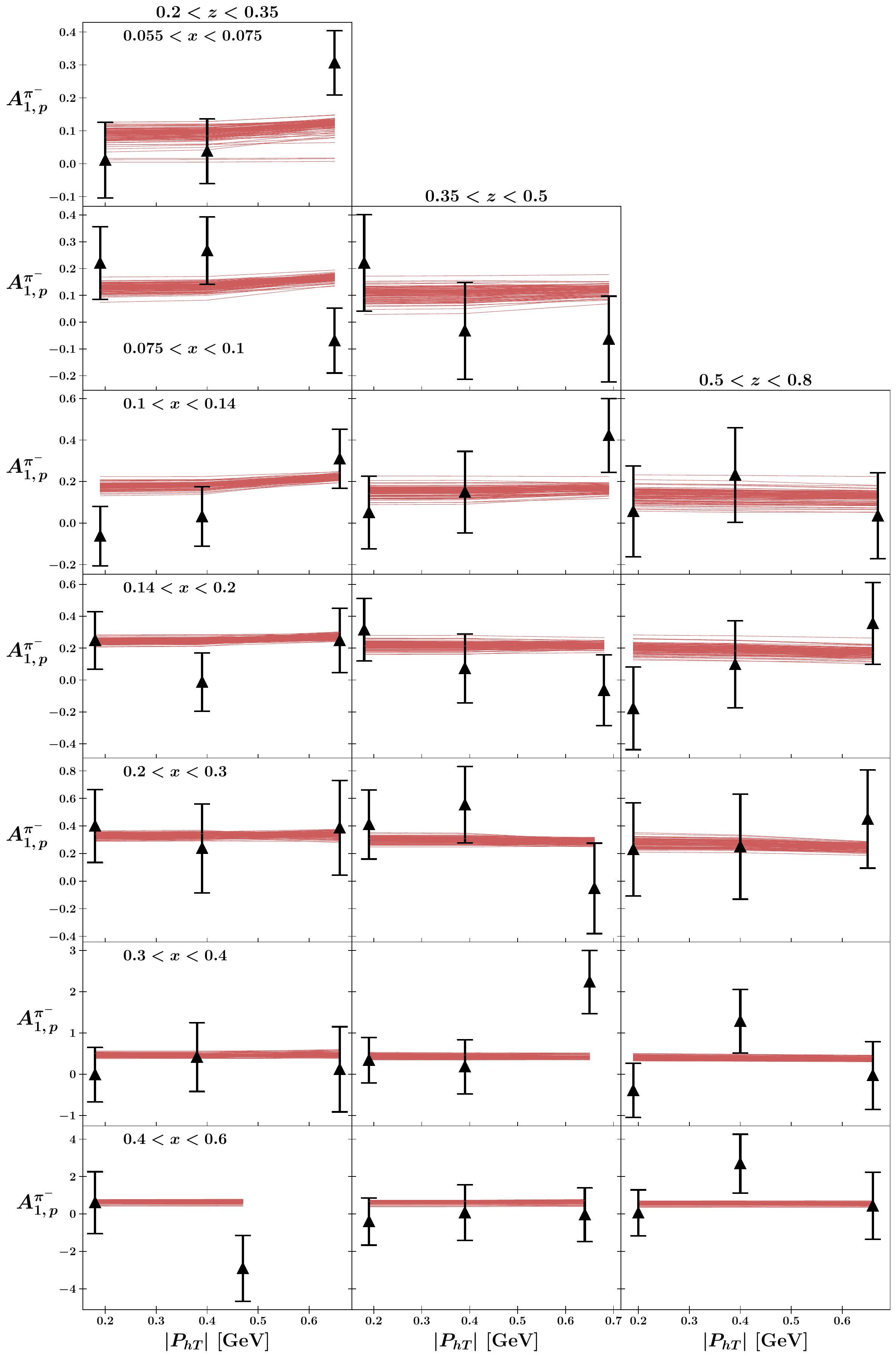}
\caption{Comparison between \textsc{HERMES} asymmetry data and theoretical predictions (red lines)  for the production of negative pion off a proton target for a selection of $x$ and $z$ bins. The results are reported as a function of the transverse momentum $|\PhT|$ of the final-state hadron.}
\label{f:A1_p_pim}
\end{figure}

\begin{figure}[h]
\centering
\includegraphics[width=0.7\textwidth]{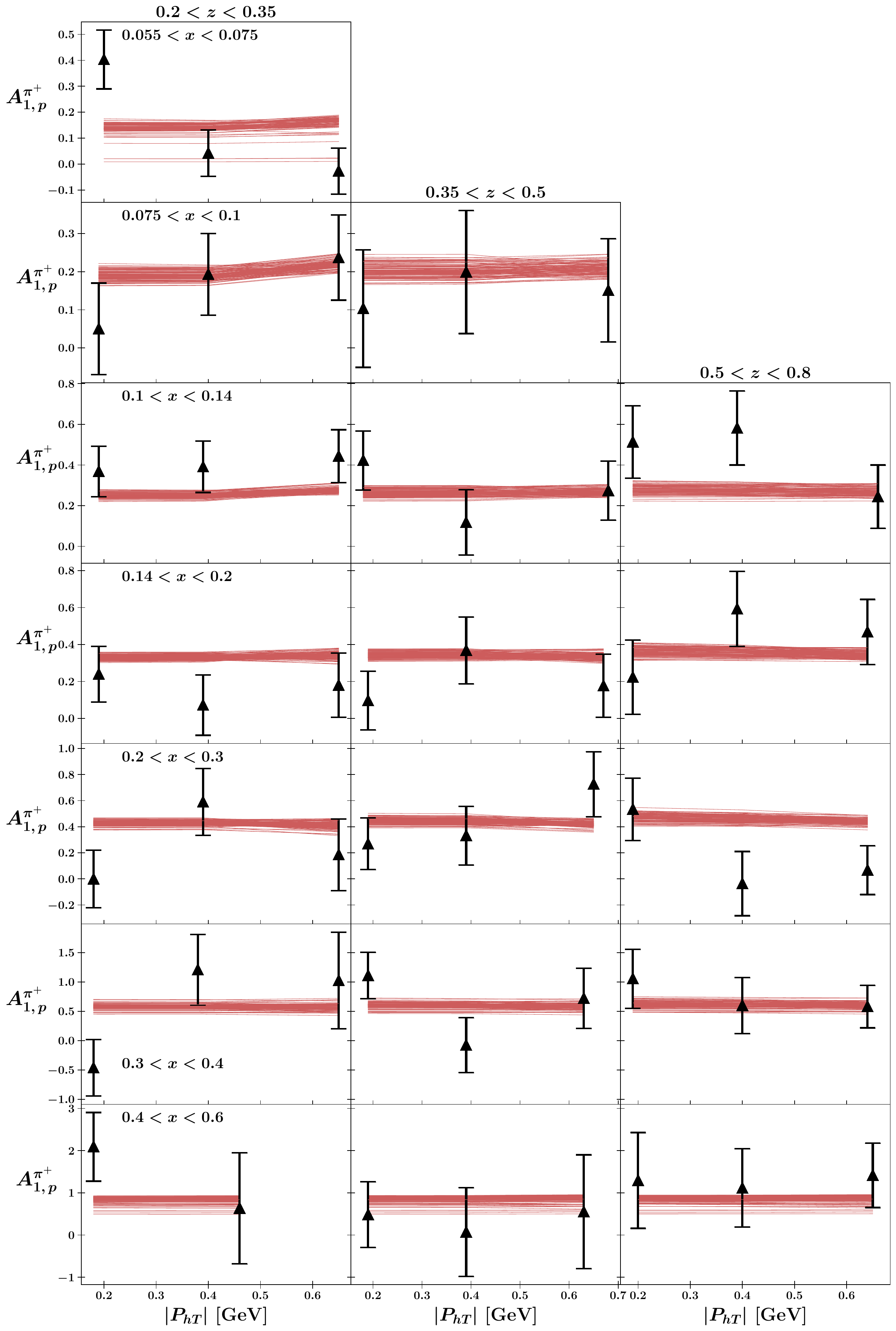}
\caption{Comparison between \textsc{HERMES} asymmetry data and theoretical predictions (red lines)  for the production of positive pion off a proton target for a selection of $x$ and $z$ bins. The results are reported as a function of the transverse momentum $|\PhT|$ of the final-state hadron.}
\label{f:A1_p_pip}
\end{figure}

\newpage

\end{widetext}

\end{document}